\documentclass[aps,prl,twocolumn,showkeys,longbibliography]{revtex4-2}
\usepackage{graphicx}
\usepackage{amsmath}
\usepackage{amssymb} 

\usepackage[english]{babel}
\usepackage{amsmath,amssymb,mathrsfs,bm,braket,color,graphicx,comment,amsfonts,mwe,tikz}
\usepackage{tabularx}
\usepackage[export]{adjustbox}
\usepackage[percent]{overpic}
 
\usepackage[colorlinks,citecolor=blue,urlcolor=blue]{hyperref}
\usepackage[mathscr]{euscript}	
\usepackage[normalem]{ulem}

\newcommand{\diag}{\text{diag}} 
\newcommand{\vect}[1]{\boldsymbol{#1}}

\begin{document} 
\title{Identifying non-Abelian anyons with upstream noise. }
\author{Misha Yutushui} 
\author{David F. Mross}
\affiliation{Department of Condensed Matter Physics, Weizmann Institute of Science, Rehovot, 76100, Israel}
\date{\today}

\begin{abstract}
Non-Abelian phases are among the most highly-sought states of matter, with those whose anyons permit universal quantum gates constituting the ultimate prize. The most promising candidate of such a phase is the fractional quantum Hall plateau at filling factors $\nu=\frac{12}{5}$, which putatively facilitates Fibonacci anyons. Experimental validation of this assertion poses a major challenge and remains elusive. We present a measurement protocol that could achieve this goal with already-demonstrated experimental techniques. Interfacing the $\nu=\frac{12}{5}$ state with any readily-available Abelian state yields a binary outcome of upstream noise or no noise. Judicious choices of the Abelian states can produce a sequence of yes--no outcomes that fingerprint the possible non-Abelian phase by ruling out its competitors. Crucially, this identification is insensitive to the precise value of the measured noise and can uniquely identify the anyon type at filling factors $\nu=\frac{12}{5}$. In addition, it can distinguish any non-Abelian candidates at half-filling in graphene and semiconductor heterostructures. 
\end{abstract}

\maketitle   

{\bf Introduction.} The fractional quantum Hall effect~\cite{Tsui_fqh_1982,Laughlin_fqh_1983,Haldane_fqh_1983,Halperin_fqh_1983,Pan_Exact_Quantization_1999} has heralded many essential concepts of modern quantum many-body physics. Its legacy includes observations of fractional charge~\cite{Picciotto_fractional_charge_1998,Dolev_observation_2008} and the notion of topological order~\cite{Wen_topological_1990,Wen_Topological_1995}. Most tantalizing is the possibility of non-Abelian excitations~\cite{Wen_Non-Abelian_1991,Moore_nonabelions_1991,Read_paired_2000,Stern_Probe_Non_Abelian_2006,nayak_non-abelian_2008,Stern_non_Abelian_2010}~\cite{Wen_Non-Abelian_1991,Moore_nonabelions_1991,Read_paired_2000,nayak_non-abelian_2008,Stern_non_Abelian_2010}. Their presence implies a topologically protected space of degenerate ground states on which braiding acts as quantum gates~\cite{Freedman_modular_functor_2000,Bonesteel_Braid_Topologies_2005,Hormozi_Topological_quantum_2007,nayak_non-abelian_2008,Stern_non_Abelian_2010}. Non-Abelian quasiparticles have long been sought for their fundamental importance and possible quantum information applications. 

The most prominent candidates for non-Abelian states arise when electrons at half-filling exhibit a quantized Hall effect. Such plateaus are well-known in the first excited Landau orbital of GaAs quantum wells~\cite{Willett_observation_1987}. Additionally, they were observed in the zeroth and third Landau orbitals of monolayer graphene~\cite{Zibrov_Even_Denominator_2018,Kim_Even_Denominator_f_wave_2019}, the zeroth orbital of bilayer graphene~\cite{Ki_bilyaer_graphene_2014,Kim_bilayer_graphene_2015,Li_bilayer_graphene_2017,Zibrov_Tunable_bilayer_graphene_2017}, and the lowest two orbitals in ZnO heterostructures~\cite{Falson_Zno_2015,Falson_Zno_2018}. Their presence is attributed to the topological superconductivity of `composite fermions', emergent excitations that are charge neutral in a half-filled Landau-level. Different pairing channels of composite fermions correspond to distinct non-Abelian phases with Ising anyons that facilitate Majorana zero modes. The most prominent candidates 
$p-ip$ (Pfaffian) and $f+if$ (anti-Pfaffian) are favored by numerics to be realized in GaAs at $\nu=\frac{5}{2}$~\cite{Morf_transition_1998,Rezayi_incompressible_2000,Peterson_Finite_Layer_Thickness_2008,Wojs_landau_level_2010,Storni_fractional_2010,Rezayi_breaking_2011,feiguin_density_2008,Feiguin_spin_2009}. 

An even richer form of non-Abelian topological order is proposed to arise at $\nu=\frac{12}{5}$ and host Fibonacci anyons~\cite{Read_Beyond_1999}. Unlike Ising anyons, they are capable of realizing a set of universal quantum gates as branding operations~\cite{Freedman_modular_functor_2000,Bonesteel_Braid_Topologies_2005,Hormozi_Topological_quantum_2007,nayak_non-abelian_2008,Stern_non_Abelian_2010}. Such excitations cannot arise in any weakly-coupled fermion system, unlike Majorana zero modes that can also occur in electronic superconductors~\cite{Sato_Majorana_Superconductors_2016}. Similarly to the half-filled case, other topological orders could also occur at $\nu=\frac{12}{5}$. They  support the simpler Ising anyons~\cite{Bonderson_hierarchy_2008} or even Abelian anyons~\cite{Haldane_fqh_1983}.

Experimentally distinguishing between different candidates remains a formidable challenge. The electric Hall conductance of any plateau is universally determined by the filling factor and cannot discriminate between competing phases. The scaling of tunneling currents with temperature or voltage has been proposed to distinguish between them~\cite{Chang_Chiral_LL_2003,Radu_Tunneling_2008,Lin_Measurements_2012}. However, there is no agreement between theoretically predicted and experimentally observed exponents, even for Abelian states~\cite{Tsiper_Formation_2001,Mandal_How_universal_2001,Mandal_relevance_2002,Wan_Universality_2005,Yang_Influence_2013}.

The thermal Hall conductance contains additional information about the topological order. In particular, it is sensitive to the anyon type. Its value $\kappa_{xy}=c_- \kappa_0$ is quantized in units of $\kappa_0=\frac{\pi^2 k_B^2}{3h} T$ and can be attributed to chiral central charge $c_-$ of \textit{all}  charged and neutral edge modes. The chiral central charge $c_-$ is the difference between the central charge of downstream and upstream moving modes~\cite{Read_paired_2000}. Edges of Abelian quantum Hall states host chiral boson modes each with central charge $c_\text{Boson}=1$. Non-Abelian Ising topological order implies an unpaired chiral Majorana fermion at the edge, whose central charge is $c_\text{Majorana}= \frac{1}{2}$. Finally, the non-Abelian phase permitting Fibonacci excitations in the bulk hosts a $\mathbb{Z}_3$ parafermion edge mode with $c_\text{Parafermion}=  \frac{4}{5}$.

Directly measuring $\kappa_{xy}$ could provide the clearest signature of a topological phase. Unfortunately, thermal Hall experiments are notoriously difficult for practical and fundamental reasons. Thermal conductance in the quantum Hall regime requires a sophisticated sample design and has only been measured in a two-terminal geometry~\cite{Jezouin_Quantum_Limit_2013,Banerjee_Observed_2017,Banerjee_observation_2018,Srivastav_Universal_quantized_2019,Dutta_Isolated_2022}. The observed quantization is orders of magnitude poorer than for charge transport. Still, recent thermal transport experiments in GaAs devices find $c_- = \frac{5}{2}$, indicating $p+ip$ pairing (PH-Pfaffian)~\cite{Banerjee_observation_2018}. Alternatively, the anti-Pfaffian phase could account for the measurements if the edge is not in complete thermal equilibrium~\cite{Simon_equilibration_2018,Feldman_comment_2018,Simon_reply_comment_2018,Feldman_equilibration_2019,Simon_equilibration_2020,Asasi_equilibration_2020}.

Noise measurements can provide a versatile tool to resolve such ambiguities about the observed thermal conductance. Their implementation is often less demanding and can operate in different transport regimes. For $\nu=\frac{5}{2}$, 
Refs.~\onlinecite{MY_Identifying_2022,Spanslatt_Noise_2019,Park_noise_2020,Manna_Full_Classification_2022,Manna_Experimentally_2023} proposed possible routes to distinguishing between the three most prominent candidate phases. The experiment carried out in Ref.~\onlinecite{Dutta_novel_2022} followed a somewhat different approach that directly focused on an emergent particle-hole symmetry of the edges to eliminate two out of three candidates. It measured upstream thermal noise at the interfaces between $\nu=\frac{5}{2}$ and $\nu=2,3$. This finding indicates particle-hole symmetry, uniquely singling out PH-Pfaffian among the candidate phases. For further theoretical analyses of the setup, see Ref.~\onlinecite{Spanslatt_Noise_2022}. 

\begin{figure}
 \centering 
 \includegraphics[width=0.95\linewidth]{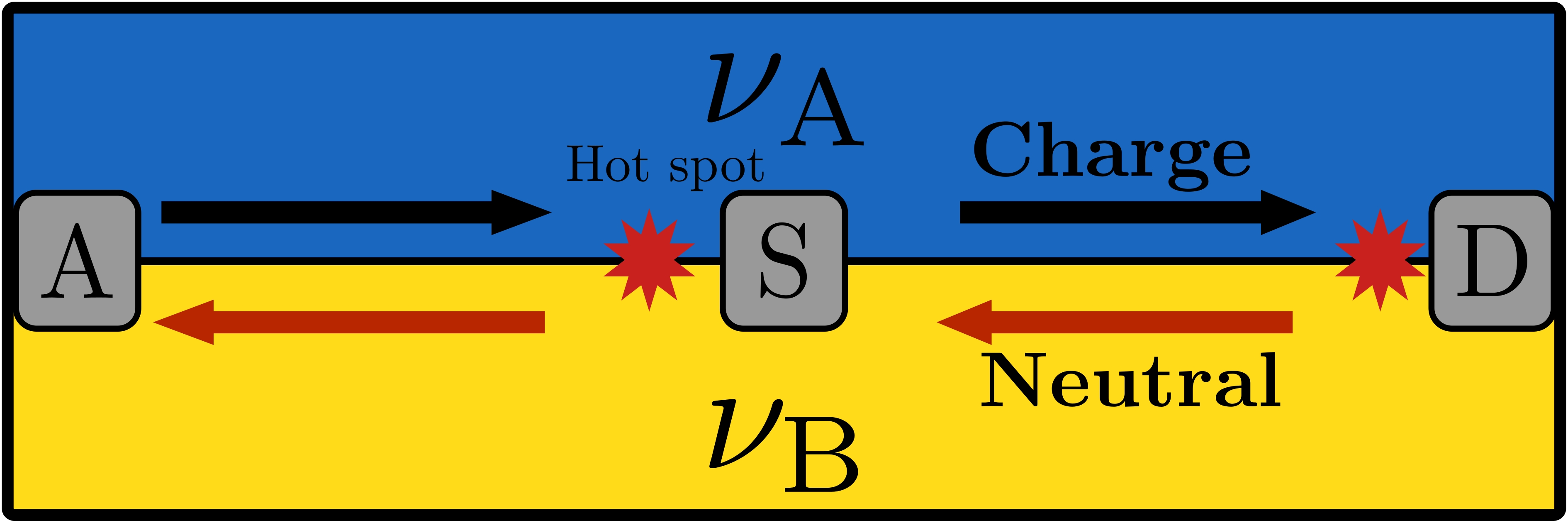}
 \caption{In a typical upstream noise measurement, a current is injected via the source (S) and flows toward the drain (D) for $\nu_\text{A}>\nu_\text{B}$. The corresponding voltage profile around the source generates a hot spot on its \textit{upstream} side, which excites neutral modes there. If one of these neutral modes propagates upstream and reaches the amplifier (A), it induces particle-hole excitations. Charge noise at the amplifier arises from a stochastic process, where either a particle or a hole flows downstream while its counterpart is absorbed. To establish well-defined upstream and downstream directions, the distance $L$ between source and drain or amplifier must be significantly larger than the scale $\ell_\text{c}$ over which charge equilibrates. The thermal equilibration length $\ell_\text{th}$ relative to the device dimension defines two distinct transport regimes. Our work primarily focuses on $L\ll \ell_\text{th}$.
}
 \label{fig.noise_setup}
\end{figure}

The typical setup of a thermal noise experiment is depicted in Fig.~\ref{fig.noise_setup}. When the distance $L$ between the source and amplifier is above the charge equilibration length, $ \ell_\text{c} \ll L$, any injected charge flows downstream into the grounded drain. Due to voltage drop, hot spots form on the \textit{upstream} sides of the source and the drain. The hot spot near the source heats any upstream neutral modes, which then propagate toward the amplifier. We further assume that the shortest thermal equilibration length between topologically protected modes is $\ell_\text{th} \gg L$. In this case, upstream neutral modes will reach the amplifier and generate noise there, even when the thermal conductance of the edge is positive (downstream). Possible mechanisms of noise generation are discussed in Refs.~\onlinecite{Park_Noise_2019,Spanslatt_Noise_2019,Park_noise_2020,Spanslatt_Noise_2022}.

The quantitative dependence of such noise on the experimental parameters is non-universal. Still, the presence or absence of noise provides direct qualitative information: 
If no upstream noise is observed, the interface is chiral. Any excess noise implies the presence of upstream neutral modes.

{\bf  Results.} To determine if the interface between two quantum Hall states, {\bf A} and {\bf B}, will be noisy, we first determine the total chiral central charge $c_-=c_\text{A}-c_\text{B}$ at the interface. Charge conservation implies at least one boson mode at the interface flowing in the downstream direction. This mode also carries energy; if no additional modes exist, then $c_- = 1$.  Conservation of charge and energy thus dictate that interfaces with $c_- < 1$ must carry upstream modes~\cite{nayak_non-abelian_2008}. The converse does not hold; interfaces with $c_- \geq 1$ may contain upstream modes and still be topologically stable (T-stable)~\cite{Haldane_Stability_1995,Moore_Classification_1997,Moore_Critical_2002}. Any interface with upstream modes exhibits noise for $\ell_\text{c} \ll L \ll \ell_\text{th}$, which is the regime we focus on. In the fully thermally equilibrated case  $\ell_\text{c}\ll\ell_\text{th}\ll L$ (at higher temperatures), noise only arises for negative $c_-$~\cite{Park_Noise_2019,Spanslatt_Noise_2019,Spanslatt_Noise_2022,Park_noise_2020}.

A general criterion for T-stability of the interface between any two Abelian phases was derived and utilized in Ref.~\onlinecite{Haldane_Stability_1995,Kao_Binding_1999}. No generalization to non-Abelian phases is known; specific cases were analyzed in Refs.~\onlinecite{Levin_particle_hole_2007,Lee_particle_hole_2007,Bishara_PH_Read_Rezayi_2008,Yang_331_MR_2017}. To investigate whether the edge is chiral or features noise, we first compute the chiral central charge. If $c_-<1$, there is noise. Otherwise, we reduce the edge to a T-stable configuration and determine if it exhibits upstream modes. 
 
We find that any non-Abelian state at half-filling can be distinguished via interfaces with the principle Haldane hierarchy states at $\nu=\frac{p}{2p+1}$ and their particle-hole conjugates at $\nu=\frac{p+1}{2p+1}$. Furthermore, the anyon type of the $\nu=\frac{12}{5}$ state can be inferred from its interfaces with $\nu=2,\frac{8}{3}$ and $3$.

{\bf Half-filled Landau level}. The candidate states for plateaus at half-integer filling factors can be considered as BCS superconductors of composite fermion. The Hall response implies that flux ${\Phi}$ is associated with a charge $Q_{\Phi} = \sigma_{xy}\Phi $, and the composite of one electron and two flux quanta is chargeless at half-filling. The pairing of these emergent neutral fermions leads to a bulk gap and topologically protected edge states. Specifically, superconductors without symmetry (Class D) are classified by an integer invariant $N \in \mathbb{Z}$. It encodes the number of chiral Majorana-fermion edge modes, i.e., the boundary of such a superconductor is described by
\begin{align}
\label{eq.Lgamma}
    {\cal L}_\gamma = i\sum_{l=1}^{|N|}\gamma_l(\partial_t - \text{sgn}(N)v_l\partial_x)\gamma_l,
\end{align}
where $\gamma_l=\gamma_l^\dag$ are Majorana fermions moving downstream (upstream) at velocity $v_l$ for $N>0$($N<0$). The physical boundary of quantum Hall states at half-filling is described by
Eq.~\eqref{eq.Lgamma} in addition to a charge mode. The latter fixes a convention for the sign of $N$, which we take to be positive for Majoranas moving in the downstream direction. As such, $N=-3$ corresponds to anti-Pfaffian~\cite{Lee_particle_hole_2007,Levin_particle_hole_2007}, $N=-1$ to PH-Pfaffian~\cite{Son_is_2015}, $N=1$ to Moore-Read~\cite{Moore_nonabelions_1991}, and 
 $N=3$ to $f$-wave pairing~\cite{Balram_parton_2018,Faugno_Prediction_2019}. 

The principle Haldane hierarchy states $\nu=\frac{p}{2p+1}$ can be viewed as integer quantum Hall states of \textit{the same} composite fermions with $\nu_\text{CF}=p$. At interfaces between integer quantum Hall states and superconductors of microscopic fermions, all counterpropagating modes localize. The resulting interfaces are fully chiral. For fractional quantum Hall states described by the analogous phases of composite fermions, the same mechanism implies a fully chiral neutral sector. Consequently, the condition $c_-=1 + \frac{N}{2}-p\geq 1$ is a sufficient criterion for the absence of upstream noise at interfaces between any hierarchy state and any candidate at half-filling. In particular, interfacing the half-filled Landau level with $\nu=\frac{1}{3}$ ($p=1$) permits a sharp distinction between $N=1$ and $N=3$.

An explicit calculation bears out the same conclusion. The interface is described by ${\cal L}_\text{edge} = {\cal L}_\gamma+ {\cal L}_K$ with~\cite{Wen_Gapless_edge_1991} 
\begin{align}\label{eq.L0}
    \mathcal{L}_K = -\frac{1}{4\pi}\sum_{a,b=0}^{p} \left[K_{ab}\partial_x\varphi_a\partial_t\varphi_b +   V_{ab}\partial_x\varphi_a\partial_x\varphi_b \right].
\end{align}
Here $K=\text{diag}(2,-K_{\text{HH}})$ is a block diagonal matrix with $K_{ab}^{\text{HH}} = 2+\delta_{ab}$~\footnote{In the operator formalism, operators $\varphi_a$ obey equal-time commutation relations $[\varphi_a(x),\varphi_b(x')] = i\pi K^{-1}_{ab}\text{sgn}(x-x')$.}. (For a discussion of edge reconstruction, see SM~\cite{Supplemental_Material}.) Edge velocities and density-density interactions are described by the non-universal matrix $V$. The coupling to electromagnetic fields is encoded in the charge vector ${\bf  t}=(1,\ldots,1)$, i.e.,  $e^{i {\bf  m} \cdot {\bf  \varphi}}$ carries electric charge $Q_{\bf  m} = {\bf  t} K^{-1} {\bf  m}$. Electrons on either side of the interface are described by $\psi _l \sim \gamma_l e^{i2\varphi_0}$ and 
$\psi_a\sim e^{i {\bf m}_a\cdot {\bf \varphi}}$, with ${\bf  m}_a = (0,K^{\text{HH}}_{a1},\ldots,K^{\text{HH}}_{ap})$ and $\mathbf{ \varphi}=(\varphi_0,\varphi_1,\ldots,\varphi_p)$.

Electron tunneling across the interface is governed by
\begin{align}\label{eq.tun}
    {\cal L}_\text{tun} &\propto \sum_{l,a}\xi_{l,a}(x)\psi^\dag_{l}\psi_{a} + \text{H.c.}, 
\end{align}
with random tunneling amplitudes $\xi_{l,a}(x)$~\cite{Kane_Randomness_1994,Moore_Classification_1997}. To analyze this interface, we introduce the total charge mode $\varphi_\text{ch}= \sum_{a}\varphi_a$ and $p$ neutral modes $\varphi'_a=2\varphi_1- {\bf  m}_a \cdot {\bf  \varphi}$. These modes decouple, i.e., commute in an operator framework. In particular, $ \psi'_a \sim e^{i \varphi'_a}$ are neutral fermions, and the electron tunneling [Eq.~\eqref{eq.tun}] is of the form $\gamma_l \psi'_a $. When tunneling is relevant, pairs of counterpropagating fermion modes localize each other, and $N-2p$ downstream Majorana fermions remain. Hence, the edge is fully chiral if $N\geq 2p$, and we predict no upstream noise.

No additional calculation is needed to understand interfaces between states at half-filling and hole-conjugate hierarchy states at $\nu=\frac{p+1}{2p+1}$. Instead, we use that this interface is identical to the one between $\nu=\frac{p}{2p+1}$ and the particle-hole conjugate of the paired state in question. The latter is again a paired state with $N \rightarrow -(N+2)$~\cite{Ken_sixteenfold_2019}, and we predict noiseless interfaces for $N \leq -2(p+1)$. Consequently, any odd-$N$ paired states are distinguishable by measuring upstream noise at the interfaces with the principal Haldane hierarchy states and their hole conjugates. (Paired states with even $N$, which describe Abelian quantum Hall states, cannot be distinguished from non-Abelian states with $N+\text{sgn}(N)$ with this procedure.)

\begin{figure}
 \centering 
 \includegraphics[width=0.95\linewidth]{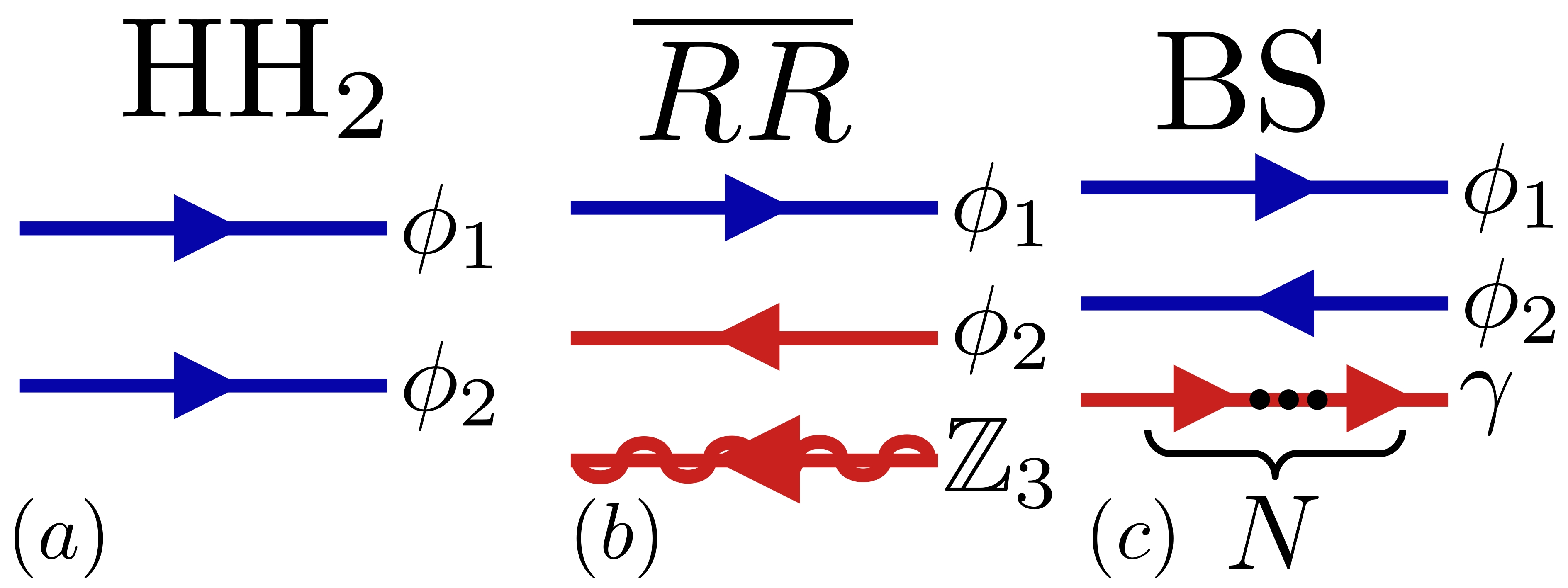}
 \caption{Three classes of candidate states for the $\nu=\frac{12}{5}$ plateau are discussed in the literature. We show their edge states here, omitting the integer modes corresponding to two filled Landau levels. {\bf (a)} The edge of the Haldane hierarchy state hosts two co-propagating boson modes. {\bf (b)} The edge of the Anti-Read-Rezayi state features a pair of counterpropagating boson modes and a $\mathbb{Z}_3$ parafermion mode.~\cite{Read_Beyond_1999,Bishara_PH_Read_Rezayi_2008} {\bf (c)} The edge of the Bonderson-Slingerland state carries $N$ downstream Majoranas and pair of counterpropagating bosons. (For $N<0$, the $|N|$ Majorana modes move upstream.) }
 \label{fig.edge_12_5}
\end{figure}

{\bf The $\nu=\frac{12}{5}$ plateau.} This filling factor can be viewed as two filled Landau levels and a partially filled one with $\nu=\frac{2}{5}$. 
The latter could support three types of candidate states with qualitatively different anyon types: Abelian, Fibonacci, and Ising. The candidate phases are:

\begin{enumerate}
    \item An Abelian hierarchy state whose edge hosts two co-propagating boson modes, i.e., $c_\text{HH}=1+1=2$.
    \item The Anti-Read-Rezayi state, which supports Fibonacci anyons in the bulk. Its interface hosts two counterpropagating boson modes and an upstream $\mathbb{Z}_3$ parafermion mode with a central charge $c_{\overline{\mathrm{RR}}}=1-1-\frac{4}{5}=-\frac{4}{5}$.
    \item The Bonderson-Slingerland state exhibits Ising topological order similar to the candidates at half-filling. It hosts two counterpropagating boson modes and $N$ downstream Majorana modes~\footnote{The Bonderson-Slingerland hierarchy states can be built on top of paired states with other pairing symmetries. The wave functions for these states can be readily written by changing the argument inside the Pfaffian similarly to Refs.~\cite{MY_Large_scale_2020,Zucker_stabilization_2016}. The resulting trial states $\Psi^\text{BS}_N=P_\text{LLL}\text{Pf}\left(\frac{1}{z}\left[\frac{\bar{z}}{z}\right]^{\frac{N-1}{2}} \right)\chi_{-2} \chi_1^3$, where $\chi_n$ is the wave function describing $n$ fully filled Landau levels, will be studied elsewhere.}. The central charge is $c_\text{BS}=1-1+\frac{N}{2}$.
\end{enumerate}

We show that the three cases are distinguishable via noise measurements at the interface between $\nu=\frac{12}{5}$ and three other states, $\nu=2,3$ and $\frac{8}{3}$ (See Table~\ref{tab.noise} for a summary). Since all candidate states include two fully filled Landau levels, the chiral central charges at the $\nu=2$ interface are $c_-=c_\alpha$ with $\alpha=\text{HH},{\overline{\mathrm{RR}}},\text{BS}$. The interfaces with $\nu=3$ each contain an additional electron mode, which reverses the flow of charge, i.e., the downstream direction. Consequently, their chiral central charges are $c_-=1- c_\alpha$. The interface with $\nu=\frac{8}{3}$ is that of $\nu=2$ supplemented by a $\nu=\frac{2}{3}$ edge~\cite{Johnson_Composite_edge_1991,Kane_Randomness_1994,Meir_Composite_1994}. The latter has a chiral central charge zero. However, the direction of charge flow is reversed, and we obtain $c_-=-c_\alpha$.

We begin with the Abelian hierarchy state. Its interface with $\nu=2$ is chiral and thus noiseless (cf.~Fig.~\ref{fig.edge_12_5}). The interface with $\nu=3$ and $\nu=\frac{8}{3}$ are characterized by $c_-=-1$ and $c_-=-2$, respectively. We thus predict noise in both cases (Table~\ref{tab.noise}).

\begin{table}[t]
    \centering
        \caption{Noise measurements on interfaces between $\nu=\frac{12}{5}$ and $\nu=2,\frac{8}{3},3$ can distinguish between the Haldane-hierarchy (HH), anti-Read-Rezayi ($\overline{\text{RR}}$), and Bonderson-Slingerland with $N=1$ (BS-MR) and $N=-3$ (BS-aPf). For each interface, we list our prediction for upstream noise and the chiral central charge.
    }
    \def\arraystretch{1.5}
    \begin{tabular}{c|| c | c | c | c} 
                        & HH \hfill ($c_-$)& $\overline{\text{RR}}$ \hfill ($c_-$) & BS-MR  \hfill ($c_-$)& BS-aPf  \hfill ($c_-$)\\\hline\hline
    $\nu=2$             & Chiral \hfill ($2$) & Noisy  \hfill ($-\frac{4}{5}$) & Noisy \hfill ($\frac{1}{2}$) & Noisy  \hfill ($-\frac{3}{2}$)\\ 
    $\nu=\frac{8}{3}$   & Noisy  \hfill ($-2$) & Noisy  \hfill ($\frac{4}{5}$)& Noisy \hfill ($-\frac{1}{2}$) & Chiral  \hfill ($\frac{3}{2}$)  \\ 
    $\nu=3$             & Noisy  \hfill ($-1$) & Chiral  \hfill ($\frac{9}{5}$)& Noisy  \hfill ($\frac{1}{2}$) & Chiral \hfill ($\frac{5}{2}$)  
    \end{tabular}
    \label{tab.noise}
\end{table} 

Interfaces of the Anti-Read-Rezayi state with $\nu=2,\frac{8}{3}$ have $c_-=-\frac{4}{5},\frac{4}{5}$. Both are below unity, and we expect upstream noise. The interface with $\nu=3$ has $c_-=\frac{9}{5}$, compatible with a chiral edge. Indeed, the Anti-Read-Rezayi state is particle-hole conjugate to the $\nu=\frac{3}{5}$ Read-Rezayi state, whose boundary to vacuum is chiral~\cite{Bishara_PH_Read_Rezayi_2008}. The latter interface is equivalent to that between Anti-Read-Rezayi and $\nu=3$; thus, we expect no noise there. 
 
Next, we study the Bonderson-Slingerland states with arbitrary $N$. For $N < 2$, the chiral central charge $c_-<1$ implies upstream noise at the $\nu=2$ interfaces. However, the converse does not hold, and there are cases with $c_- \geq 1$ and protected upstream modes (see SM~\cite{Supplemental_Material}). Specifically, we find that four is the smallest value of $N$ for which the interface is chiral.

The quantum numbers for $N=4$ match those of the Haldane hierarchy state, suggesting they represent the same phase. We analyze the interface between the $N=4$ Bonderson-Slingerland state and $\nu=2$ to test this possibility. It is described by
${\cal L}_\text{edge} = {\cal L}_\gamma+ {\cal L}_{K}$ [cf.~Eqs.~(\ref{eq.Lgamma},\ref{eq.L0})] with $K_\text{BS} = \begin{pmatrix}2 & 3 \\ 3 & 2\end{pmatrix}$ and ${\bf  t}=\begin{pmatrix}1 \\ 1 \end{pmatrix}$. We combine two downstream Majorana modes according to $\gamma_1 + i \gamma_2 \sim e^{i \varphi_3}$. Next, we introduce two \textit{upstream} Majoranas $\eta_1 + i \eta_2 \sim e^{i(\varphi_1 - \varphi_2 + \varphi_3)}$. The latter can localize the remaining downstream Majoranas $\gamma_{3,4}$ through the tunneling of electrons $\psi_{e1} \sim \gamma_l e^{i(2\varphi_1+3\varphi_2)}$ or $ \psi_{e2}\sim\gamma_l e^{i(3\varphi_1+2\varphi_2)}$, i.e.,  
\begin{align}
    {\cal L}_\text{tun} \propto \xi(x) \psi^\dag_{e1}\psi_{e2}+\text{H.c.}\sim i \xi(x) \gamma_l \eta_{l'}+\text{H.c.}
\end{align}
The modes $\varphi'_{1,2} = \varphi_{1,2}\pm \varphi_3$ decouple from the tunneling term, i.e., commute with its phase in an operator framework. They consequently persist and are described by ${\bf  t}' = {\bf  t}$ and $K'=K_\text{HH}$. It follows that Bonderson-Slingerland with $N=4$ is indeed equivalent to the $\nu=\frac{2}{5}$ hierarchy state and thus fully chiral.

More generally, the Bonderson-Slingerland state with $N$ downstream Majoranas corresponds to hierarchy states with $N-4$ downstream Majoranas. As such, we predict upstream noise at the $\nu=2$ interface for $N \leq 3$. For a detailed calculation, see SM~\cite{Supplemental_Material}.

For $\nu=3$, we similarly find that central charge considerations alone are insufficient, and $N\leq -2$ is required for a chiral edge. This interface differs from the previous one by a chiral electron mode $\psi_{e0} \sim e^{i \varphi_0}$ moving oppositely to the $\frac{2}{5}$ fractional charge modes. Electron tunneling between integer and fractional edges is given by
\begin{align}
    {\cal L}_\text{tun} \propto \xi(x) \psi^\dag_{e0}\psi_{e1}+\text{H.c.}=\xi(x) \gamma_l\tilde\eta_{l'}+\text{H.c.},
\end{align} 
where $\tilde\eta_1 + i \tilde\eta_2 \sim e^{i(\varphi_0+2\varphi_1+3\varphi_2)}$ defines two upstream Majorana fermions. As in the case of half-filling, we thus have a total of $-(N+2)$ downstream Majorana modes at the hole-conjugate edge. The modes $\varphi_1'' = \varphi_0+\varphi_1+\varphi_2$ and $\varphi_2'' =-\varphi_0-\varphi_1-2\varphi_2$ decouple as above. They are described by $K''=K_\text{HH}$ and charge vector ${\bf  t}''=(1,0)$. Consequently, we predict upstream noise for $N \geq -1$.

Finally, for interfaces with $\nu=\frac{8}{3}$, we add an upstream $\nu=\frac{1}{3}$ mode $\varphi''_0$ to the hole-conjugate edge that we obtained in the previous paragraph~\footnote{ Several microscopic edge structures have been proposed theoretically~\cite{Johnson_Composite_edge_1991,Kane_Randomness_1994,Meir_Composite_1994}. Under our assumption that only topologically protected modes survive beyond $\ell_c$, all of these lead to the same conclusion.}. Following the analysis of Ref.~\cite{Haldane_Stability_1995}, the composite edge is T-unstable; electron tunneling between these edges ${\cal L}_\text{tun}\sim \cos(3\varphi''_0+3\varphi''_1+2\varphi''_2)$ localizes a pair of counterpropagating modes. The reduced edge hosts a single $\nu=\frac{4}{15}$ charge mode and $-(N+2)$ Majoranas. The resulting condition of $N\leq -2$ for a chiral edge saturates the $c_-\geq 1$ requirement.

{\bf  Discussion.} We have demonstrated that upstream noise is a powerful and versatile tool for identifying non-Abelian quantum Hall states. Interfacing the plateau under investigation with a carefully selected Abelian phase yields a qualitative outcome of noise or no noise that differentiates between two candidate states. By process of elimination, the topological order can thus be identified. Crucially, such an experiment does not rely on the precise value of the noise and can be interpreted without having to assume any particular noise-generating mechanism.

For the half-filled Landau level, we found that any odd pairing channel can be distinguished by interfacing with Haldane hierarchy states at $\nu=\frac{p}{2p+1}$ and their particle-hole conjugates at $\nu=\frac{p+1}{2p+1}$ (on top of the same number of filled Landau levels). The interfaces are chiral if the number of Majoranas $N\geq 2p$; for hole-conjugate hierarchy states, the condition is $N\leq -(2p+2)$. For example, an interface with $\nu=\frac{1}{3}$ differentiated between Moore-Read Pfaffian and $f$-wave pairing proposed in Ref.~\onlinecite{Kim_Even_Denominator_f_wave_2019}.

For the case of $\nu=\frac{12}{5}$, interfaces with the nearby integers $\nu=2,3$ suffice to differentiate between the three proposed candidate phases: A Haldane hierarchy state, the anti-Read-Rezayi state, and the Bonderson-Slingerland state with $N=1$ (Moore-Read pairing). Measuring the interface with $\nu=\frac{8}{3}$ as well can further distinguish these phases from the $N=-3$ generalization of the Bonderson-Slingerland state (anti-Pfaffian pairing). 

In any quantum Hall system, edge reconstruction may occur and mask the bulk phase. In particular, topologically unprotected modes may be present near the edge and hybridize with those dictated by the topological state of the bulk. Such spurious modes can strongly modify short-distance properties, i.e., near a quantum point contact, but not long-distance physics. Our analysis assumes that $\ell_\text{c}$ sets the scale where unprotected modes drop out, and we require $L \gg \ell_\text{c}$. 

Finally, we want to contrast noise experiments with thermal edge conductance measurements. The latter requires full equilibration of charge and heat, $L \gg \ell_\text{c},\ell_\text{th}$, to reveal the bulk topological order. Experiments typically operate in the regime where $\ell_\text{th} \gg \ell_\text{c}$, especially at the lowest temperatures~\cite{Banerjee_Observed_2017,Banerjee_observation_2018,Dutta_novel_2022}.The experimental data of Ref.~\cite{Dutta_novel_2022} is only consistent with the regime where $\ell_\text{th} \gg L \gg \ell_\text{c}$. If instead $L \gg \ell_\text{th},\ell_\text{c}$ either the interface with $\nu=2$ or with $\nu=3$ would be noiseless \textit{independent of the $\nu=5/2$ state}.

\noindent{\bf \large Acknowledgments}\\
It is a pleasure to thank Ankur Das, Sourav Manna, Yuval Gefen, Bivas Dutta, Yuval Ronen and Moty Heiblum, for illuminating discussions on this topic. This work was partially supported through CRC/Transregio 183, and the Israel Science Foundation (2572/21).

\bibliography{phpfbib}

\appendix

\section{Appendix: Edge reconstruction}
\label{app.A}
The bulk topological order of a quantum Hall phase dictates certain properties of the edge, such as charge and thermal conductance. In real systems, the edge may have an internal structure where bulk filling factor $\nu_\text{bulk}$ is separated from vacuum by a thin strip of a different filling factor $\nu_\text{strip}$. The composite edge is then comprised of $\nu_\text{bulk}$ to $\nu_\text{strip}$ interface modes and $\nu_\text{strip}$ to vacuum edge modes. As such, it may (but need not) contain more modes than are demanded by the bulk topology. Such modes may disguise the universal edge physics at short scales. Other ways of localizing excess modes may realize multiple distinct edge phases, cf.~Ref.~\onlinecite{Cano_Bulk_edge_2014}.

\subsection{ Hierarchy state at $\nu=\frac{2}{5}$}
The edge of the Haldane hierarchy state at $\nu_\text{bulk}=\frac{2}{5}$ is typically described by one of two edge theories. The first theory is given by Eq.~(2) in the main text with the same K-matrix and charge vector that encode the bulk topological order, i.e., $K_1 = \begin{pmatrix}3 & 2 \\ 2 & 3\end{pmatrix}$ and $t_1=\begin{pmatrix}1 \\ 1\end{pmatrix}$. Edge electrons are $\psi_a \sim e^{i \Phi_a}$ with $\vect \Phi =K\vect \varphi$, and elementary charge $\frac{e}{5}$ quasiparticles are $\chi_a \sim e^{i \varphi_a}$. 

\begin{figure}[h]
 \centering 
 \includegraphics[width=0.8\linewidth]{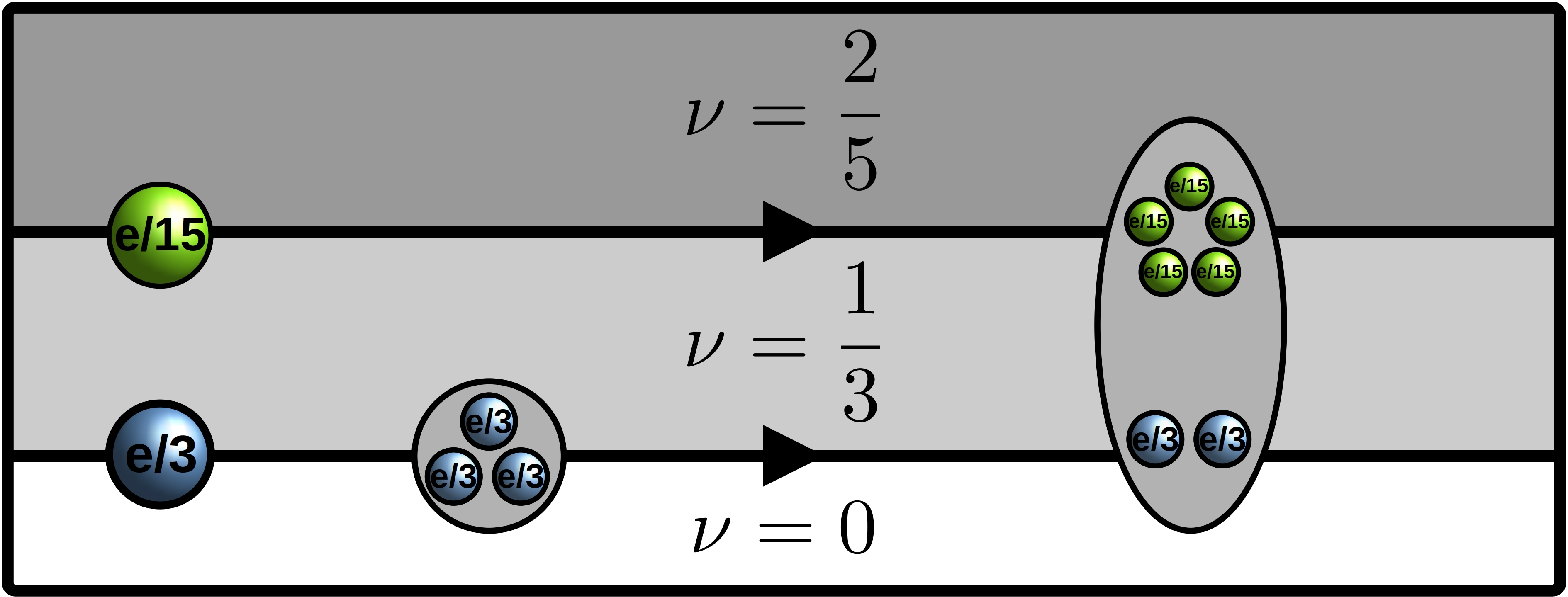}
 \caption{ The edge structure of the $\nu=\frac{2}{5}$ hierarchy state is separated from the vacuum by a thin $\nu=\frac{1}{3}$ strip. The two most relevant electron operators $e^{ei\varphi_1}$ and $e^{i(2\varphi_1+5\varphi_2)}$ are shown. The one on the right is an example of $\frac{e}{3}$ quasiparticle-quasihole dressing of the left electron that consists of three $\frac{e}{3}$ quasiparticles.} \label{fig.strips}
\end{figure}

The second theory assumes an intermediate $\nu_\text{strip}=\frac{1}{3}$ strip at the boundary, as shown in Figure~\ref{fig.strips}. The interface between strip and bulk hosts a charge mode that accounts for the difference of filling factors $\nu_\text{bulk}-\nu_\text{strip}=\frac{1}{15}$. As a result, the composite edge is described by $K_2 = \begin{pmatrix}3 & 0 \\ 0 & 15\end{pmatrix}$ and $t_2=\begin{pmatrix}1 \\ 1\end{pmatrix}$. Tunneling of charge $\frac{e}{3}$ quasiparticles is described by $e^{i (\varphi_1 - 5 \varphi_2)}$. Any edge electron can be obtained by dressing $e^{3 i \varphi_1 }$ with such processes, i.e., $\psi \sim e^{3 i \varphi_1 +i n (\varphi_1 - 5 \varphi_2)}$. Elementary quasiparticles that are mutually local with all electrons are $\chi_1 \sim e^{i \varphi_1 -2i\varphi_2}$ and $\chi_2 \sim e^{3 i \varphi_2}$; both carry $\frac{e}{5}$ charge.

To show that the two descriptions of the $\nu=\frac{2}{5}$ edge are stably equivalent, we supplement an edge described by $K_1$ with a topologically trivial pair of counterpropagating $\nu=\pm \frac{1}{3}$ modes, i.e., $K =\text{diag}(K_1,-3,3)$ with $t_a=1$. Such modes can be viewed as a narrow $\nu=\frac{1}{3}$ strip placed near the $K_1$ edge, separated from the bulk by a small vacuum strip. When tunneling of $\frac{e}{3}$ \textit{quasiparticles} via
\begin{align}
\label{eqn.qptunneling}
    {\cal L}_1 \propto \xi_1(x)e^{i(\varphi_3+\varphi_4)}+\text{H.c.}
\end{align}
is relevant, it localizes the additional modes and recovers the $K_1$ edge. Alternatively, \textit{electron} tunneling across the vacuum strip via
\begin{align}
\label{eqn.etunneling}
    {\cal L}_2 \propto \xi_2(x)e^{i(2\varphi_1+3\varphi_2 + 3\varphi_3)}+\text{H.c.}
\end{align}
localizes a different pair of counterpropagating modes. This process leaves behind a $\nu=\frac{1}{15}$ mode and the outer $\nu=\frac{1}{3}$ mode, i.e., the edge theory $K_2$. 

To capture both cases, we introduce the basis $\varphi_1'=\varphi_1$, $\varphi_2'=\varphi_2 + \varphi_3 + \varphi_4$, $\theta = \varphi_3+\varphi_4$, and $2\phi = 2\varphi_1 +3\varphi_2 + 3\varphi_3$. The edge theory for $(\varphi_1',\varphi_2',\theta,\varphi)$ is described by $K=\diag(K_1,-2\sigma_x)$ and charge vector $\vect t=(1,1,0,0)$. In particular, the non-chiral fields $\theta$ and $\varphi$ are canonical conjugates and decouple from  $\varphi'_{1,2}$. In this basis, the two tunneling terms become
\begin{align}\label{eq.cosines}
    {\cal L}_1\sim \xi_1(x)\cos[\theta]~,\qquad
    {\cal L}_2\sim \xi_2(x)\cos[2\phi]~.
\end{align}
Sine-Gordon models with competing cosines have been extensively studied in cf.~Ref.~\onlinecite{Lecheminant_Criticality_2002}. In particular, it is known that no phase transition occurs between limits where ${\cal L}_1$ or ${\cal L}_2$ dominate. We note that if only electron tunneling across the $\nu=\frac{1}{3}$ strip were possible, the cosine in ${\cal L}_1$ would be replaced by $\cos[ 3\theta]$. In that case, there would be a phase transition.

One immediate consequence of the equivalence between these edges is that both descriptions yield the same scaling of tunneling currents with voltage and temperature. The most relevant electron operators, the minimal charge quasiparticle operators, and the most relevant quasiparticle operators have the scaling dimensions $\Delta_e=\frac{3}{2}$, $\Delta_\frac{e}{5} = \frac{3}{5}$, and $\Delta_\frac{2e}{5} = \frac{2}{5}$, respectively.

\subsection{ Hierarchy state at $\nu=\frac{p}{2p+1}$}
We now consider a Haldane-hierarchy at filling factor $\nu_\text{bulk}=\frac{p}{2p+1}$ and K-matrix $K_{ab} = 2+\delta_{ab}$. Edge electrons are $\psi^p_a\sim e^{i \Phi^\text{el}_a}$ with $\vect \Phi_\text{el}= K \vect \varphi$; elementary quasiparticles with charge $\frac{e}{2p+1}$ are $\chi_a^p\sim e^{i\varphi_a}$. In an operator formalism, they satisfy
\begin{align}
\label{eq.eqpcomm}
[\Phi^\text{el}_a, \varphi_b] =i\pi \delta_{ab}~.
\end{align}
The bulk phase may be separated from the vacuum by a narrow strip of lower filling factor $\nu_\text{strip}=\frac{q}{2q+1}$. We denote electrons (quasiparticles) on the downstream and upstream sides of the strip by $\psi^q$ ($\chi^q$) and $\bar\psi^q$ ($\bar\chi^q$), respectively. As such, tunneling of charge $\frac{e}{2q+1}$ quasiparticles across the strip is 
\begin{align}
{\cal L}_1 \sim \sum_{a=1}^q \bar \chi^{q\dagger}_a \chi^{q}_a \sim \sum_{a=1}^q\cos \theta_a~.
\end{align}
generalizing Eq.~\eqref{eqn.qptunneling}. Electron tunneling between $p$ and $q$ edges is
\begin{align}
{\cal L}_2\sim \sum_{a=1}^q \bar \psi^{q\dagger}_a \psi^{p}_a \sim \sum_{a=1}^q\cos 2\phi_a~.
\end{align}

Eq.~\eqref{eq.eqpcomm} implies that $\theta_a,\phi_a$ implicitly defined by ${\cal L}_{1,2}$ are canonical conjugates. It follows that there is no phase transition between the limits where ${\cal L}_{1}$ or ${\cal L}_{2}$ dominate. More generally, any sequence of strips with successively lower filling factors is equivalent in the same sense.

\section{ Appendix: Bonderson-Slingerland state}\label{app.2}

Bonderson and Slingerland introduced~\cite{Bonderson_hierarchy_2008} the eponymous state in a different basis than we use here. To facilitate connections to their work, we summarize the quasiparticle content of this phase in our basis.
We express the bosonic sector of the theory in the `symmetric' basis with $K_\text{BS} =\begin{pmatrix}2 & 3 \\ 3 & 2\end{pmatrix}$ and $\vect t=\begin{pmatrix}1 \\ 1\end{pmatrix}$. The same K-matrix and charge vector also describe a \textit{bosonic} $\nu=\frac{2}{5}$ state. In addition, the Bonderson-Slingerland state contains a neutral fermion sector whose edge manifestation is a chiral downstream Majorana fermion. We permit an arbitrary number $N$ of Majorana fermions $\gamma_l$. The local edge electrons are combinations of a Majorana fermion and charge-one quasiparticles of the bosonic sector $\gamma_l e^{i\vect m \cdot \vect \varphi}$ with $\vect{m}=(3,2)$ or $\vect{m}=(2,3)$. 

When $N=2n$ or $N=2n+1$ with $|n|>0$, we combine $n$ pairs of Majorana fermions $\gamma_{2k-1}+i\gamma_{2k}\sim e^{i\varphi_{k}}$ with $k=1,\ldots,n$. Then, the edge is described by $K=\diag(K_\text{BS},s,\ldots,s)$ with $s=\text{sgn}(n)$ corresponding to the chirality of Majoranas. When $N$ is even, the quasiparticles 
\begin{align}\label{eq.even}
     {\cal O}^\mathbb{I}_{\vect m}\sim e^{i \vect m \cdot \vect \varphi},\quad
    {\cal O}^{\sigma}_{\vect m}\sim e^{i (\vect m +\frac{1}{2}) \cdot \vect \varphi}
\end{align}
are mutually local with electrons when $m_a\in \mathbb{Z}$~\cite{Ken_sixteenfold_2019}. If $N=2n+1$, there is one unpaired Majorana $\gamma\equiv\gamma_N$. In this case, the quasiparticles mutually local with electrons are
\begin{align}\label{eq.odd}
    {\cal O}^\mathbb{I}_{\vect m}\sim e^{i \vect m \cdot \vect \varphi},\quad
    {\cal O}^{\gamma}_{\vect m}\sim \gamma e^{i \vect m \cdot \vect \varphi},\quad
    {\cal O}^{\sigma}_{\vect m}\sim \sigma e^{i (\vect m+\frac{1}{2}) \cdot \vect \varphi}.
\end{align}
For either even or odd $N$, the minimal non-zero charge of a quasiparticle is $\frac{e}{5}$ in all sectors.

\subsection{The interface with $\nu=2$}\label{app.BS_2} 
In the main text, we showed that the interface with $\nu=2$ is chiral when the number of Majoranas $N\geq4$. Meanwhile, the edges with $N<2$ are non-chiral based on chiral central charge considerations. Here, we show that edges with $N=2$ and $N=3$ are also non-chiral. 

The $N=2$ case is Abelian, and its T-stability can be determined as in Ref.~\onlinecite{Haldane_Stability_1995}. Specifically, we have $K=\diag (K_{BS},1)$ and $\vect t = (1,1,0)$. The edge is T-unstable if there exists a non-trivial operator $\mathcal{O}_{\vect m} = e^{i\vect m\cdot \vect \varphi}$ that (i) consists of bulk quasiparticle [Eq.~\eqref{eq.even}], (ii) is charge neutral $\vect t K^{-1}\vect m=0$, and (iii) has zero topological spin $h_{\vect m}=\frac{1}{2}\vect m K^{-1}\vect m=0$~\footnote{The phase accumulated by the clockwise exchange of identical quasiparticles $\alpha$ is given by a topological spin $e^{2i\pi h_\alpha}$.}. The first condition implies that $\vect{m}$ can either have all integer or all half-odd-integer entries. The second condition implies $m_1 = -m_2$. The final condition requires $2m_1^2 = m_3^2$ and has no integer solutions. This implies that the edge is T-stable.

In the $N=3$ case, the edge is T-unstable. In the basis $\varphi'_{1,2}=\varphi_{1,2}\pm \varphi_3$ and $\varphi'_3 = \varphi_1-\varphi_2+\varphi_3$, we have $K' = \diag(K_\text{HH},-1)$. The unpaired downstream Majorana $\gamma$ localizes with one of the Majoranas $\eta_1+i\eta_2\propto e^{i\varphi_3'}$, leaving behind an unpaired {\it upstream} Majorana. The quasiparticle spectrum is given by Eq.~\eqref{eq.odd} with $\varphi\to \varphi'$ and $\gamma\to \eta$. 

As this edge is still achiral, the question about its T-stability arises. In this case, we need to apply conditions (i)-(iii) to the operators $\mathcal{O}^{\mathbb{I}}_{\vect m}$ and $\mathcal{O}^\eta_{\vect m}$. Similar to the $N=2$ case, charge conservation implies $m_1=-m_2$. The conformal spins of the two operators are then $h_{\mathbb{I}}=m_1^2$ and $h_{\eta}=m_1^2- \frac{1}{2}$, respectively. There is no non-zero integer solution, and we conclude that the $N=3$ state also exhibits noise at the $\nu=2$ interface. 

\subsection{ The interface with $\nu=3$}
In the main text, we showed that  the edge is chiral if $N\leq-2$. The central charge ensures that the edge exhibits noise for $N\geq1$. Here we cover the two remaining cases, and we show that the $N=0$ and $N=-1$ edges are non-chiral. 

The $N=0$ case is described by $K=\diag(1,-K_\text{BS})$ and $\vect t'=(1,1,1)$ and can be analyzed within the framework of Ref.~\onlinecite{Haldane_Stability_1995}. The equations (i)-(iii) have no non-zero, integer solution, which implies that the edge is T-stable and contains an upstream mode. 

The $N=-1$ case is T-unstable, similar to the $N=3$ interface with $\nu=2$. In the basis $\varphi_0''=\varphi_0+2\varphi_1+3\varphi_2$, $\varphi_1'' = \varphi_0+\varphi_1+\varphi_2$, and $\varphi_2''=-\varphi_0-\varphi_1-2\varphi_2$, the interface action is Eq.~(2) in the main text with $K''=\diag(-1,K_\text{HH})$ and $\vect t''=(0,1,0)$. One of the two Majoranas $\tilde{\eta}_1+i\tilde{\eta}_2\sim e^{i\varphi_0'}$ gets localized with $\gamma$. Therefore, the resulting theory is described by $K_\text{HH}$ with $\vect t = (1,0)$ in addition to one upstream Majorana $\eta$. 

Its T-stability is verified using the same conditions as before. Charge conservation demands $3m_1=2m_2$. Then, the smallest topological spin of the charge-neutral operator $O_{\vect m}=e^{i{\vect m} \cdot {\vect \varphi}}$ is $h_{\vect m}=\frac{3}{2}$ for ${\vect m}=(2,3)$. It cannot be compensated for by $\eta$ with $h_{\eta}=-\frac{1}{2}$. Therefore, the edge is T-stable and exhibits noise due to the upstream Majorana.

\end{document}